\documentclass[journal]{IEEEtran}
\usepackage{graphicx}

\usepackage{subcaption}
\usepackage{hyperref}
\usepackage[labelformat=simple]{subcaption}

\usepackage{float}
\usepackage{tabularx}
\usepackage{booktabs}
\usepackage{multirow}
\usepackage{amsmath,amssymb}
\usepackage{mathtools}

\usepackage[dvipsnames]{xcolor}

\newcommand*{\expec}{\mathbb{E}}

\DeclareMathOperator{\power}{P}
\DeclareMathOperator{\RM}{RM}

\DeclareMathOperator{\dBm}{dBm}
\DeclareMathOperator{\GHz}{GHz}

\DeclareMathOperator{\Tx}{T}
\DeclareMathOperator{\Rx}{R}

\newlength{\imgwidth}
\title{Radio Map Prediction from Noisy Environment Information and Sparse Observations}

\author{Fabian Jaensch, 
\c{C}a{\u g}kan Yapar,
Giuseppe Caire,
Beg\"{u}m Demir 
\thanks{F. Jaensch, \c{C}. Yapar and G. Caire are with the Communications and Information Theory Group, Technische Universit{\"a}t Berlin, 10623 Berlin, Germany.}
\thanks{B. Demir is with the Remote Sensing Image Analysis Group, Technische Universit{\"a}t Berlin, 10623 Berlin, Germany, and with the BIFOLD - Berlin Institute for the Foundations of Learning and Data, 10623 Berlin, Germany.}
\thanks{Corresponding author: Fabian Jaensch (email: f.jaensch@tu-berlin.de).}
}

\begin{document}
\maketitle


\begin{abstract}
    Many works have investigated radio map and path loss prediction in wireless networks using deep learning, in particular using convolutional neural networks.
    However, most assume perfect environment information, which is unrealistic in practice due to sensor limitations, mapping errors, and temporal changes. 
    We demonstrate that convolutional neural networks trained with task-specific perturbations of geometry, materials, and Tx positions can implicitly compensate for prediction errors caused by inaccurate environment inputs. 
    When tested with noisy inputs on synthetic indoor scenarios, models trained with perturbed environment data reduce error by up to 25\% compared to models trained on clean data.
    We verify our approach on real-world measurements, achieving 2.1 dB RMSE with noisy input data and 1.3 dB with complete information, compared to 2.3-3.1 dB for classical methods such as ray-tracing and radial basis function interpolation. 
    Additionally, we compare different ways of encoding environment information at varying levels of detail and we find that, in the considered single-room indoor scenarios, binary occupancy encoding performs at least as well as detailed material property information, simplifying practical deployment.
    \begin{IEEEkeywords}
        Convolutional Neural Networks, Machine Learning, Path loss, Radio map, RSSI.
    \end{IEEEkeywords}
\end{abstract}

\section{Introduction}\label{sec:introduction}

Radio maps represent the spatial distribution of quantities of interest in wireless systems such as the path loss.
They are a useful tool for wireless network planning, optimization and real-time applications, e.g. localization and beam management \cite{planning}, \cite{overview},  \cite{locunet}, \cite{beammanagement}.
Statistical models that determine the path loss based solely on the distance to the transmitter (Tx) are inherently unable to capture radio wave propagation effects such as shadowing, reflections or diffractions in specific environments.
Deterministic simulations via ray-tracing (RT), on the other hand, can take a significant amount of time making them unsuitable for real-time applications or optimization tasks requiring the computation of many radio maps.
Furthermore, they require a precise representation of the environment's geometry and typically also of characteristics such as material properties, which may not always be available.

Recent advances in deep learning, particularly convolutional neural networks (CNNs), have shown promising results for radio map prediction in terms of accuracy, speed and the flexibility to work with different kinds of inputs.
Seminal works have shown that CNNs trained on large datasets of synthetic radio maps produced by RT in outdoor scenarios are capable of accurately predicting radio maps from given environment information and Tx locations \cite{radiounet}, \cite{fadenet}, \cite{plnet}.
Later works have explored adaptation of the approach to indoor scenarios as well \cite{deepray}, \cite{challengeindoor}.
Different kinds of inputs such as images instead of explicit environment geometry have been considered in \cite{plgan}, \cite{imgopt}. 
Additional information (e.g. antenna patterns) has been used in \cite{plnet}, \cite{challengeindoor}, \cite{imgopt}, and physic-inspired inputs such as free space path loss or electromagnetic properties of objects in \cite{deepray}, \cite{challengeindoor}.

In this work, we study radio map prediction under the practically relevant assumption that environment information is incomplete and inaccurate, while a limited number of signal strength observations from the ground truth are available. 
We treat environment data as an uncertain prior and sparse observations as reliable but spatially limited anchors. 
Our approach trains CNNs with physically motivated perturbations of geometry, materials, and transmitter locations to improve robustness to such inaccuracies. 
We further investigate how much detail in the environment encoding is actually beneficial under these conditions.

In \cite{spatial_freq}, a tensor of radio maps at different frequencies is completed from measurements in different spatial locations and at different frequencies.
A special CNN architecture featuring 1D convolutions to mix features from different frequencies and 2D convolutions applied to features from different spatial locations separately for each frequency is shown to outperform other architectures directly mixing information from different locations and frequencies with 2D convolutions.
In the competition \cite{challengeindoorsamples}, participants were asked to develop ML-based methods incorporating sparse ground truth samples to predict indoor radio maps.
In addition to finding suitable architectures and input encodings to achieve low error given random samples, a second task in the challenge consisted of determining optimal sampling positions to guide the radio map prediction.

In machine learning and computer vision, while most research assumes clean training data, several works have also investigated robustness to input noise and label corruption.
As an example, in \cite{noise_med} it is demonstrated that training object recognition networks with strong pixel noise improves robustness and better matches human vision characteristics.
Similarly, \cite{robust_adversarial} introduces stochastic noise models applied to inputs and intermediate layers, showing improved generalization on image classification benchmarks.
For label noise, \cite{labelnoise} provides a comprehensive review of techniques including robust loss functions, sample reweighting, and iterative label correction methods that improve accuracy on noisy medical datasets.

This body of work mainly considers unstructured or task-agnostic noise, such as pixel corruption or label flips. 
In contrast, inaccuracies in radio map prediction arise from structured, physically meaningful perturbations of geometry, materials, and transmitter locations. 
Such perturbations directly affect wave propagation and cannot be adequately modeled by generic stochastic noise.

Most existing CNN-based approaches for radio map prediction further assume that complete and accurate environment information is available.
In our previous work \cite{imgopt}, we relaxed this assumption and investigated radio map prediction in outdoor scenarios from incomplete environment information in the form of aerial images.
In this work, we shift the focus to indoor environments, where environment representations are not only incomplete but also affected by systematic inaccuracies.

In practice, indoor 3D environment models may be incomplete or outdated as furniture or other objects move or temporary obstacles appear.
Sensors such as LiDAR or cameras may fail to capture parts of the environment due to occlusions, and environment models reconstructed from their measurements, such as point clouds or images, inevitably contain geometric inaccuracies.
Material properties are often unknown, and even when environment models exist, object and transmitter positions may contain errors due to mapping inaccuracies or temporal changes.
While physics-based tools such as differentiable RT frameworks \cite{sionna_rt} allow fitting certain parameters (e.g., material properties), they remain highly sensitive to geometric inaccuracies and therefore still require precise environment geometry.

At the same time, acquiring a limited number of signal strength measurements (\textit{observations}\footnote{We use the term \textit{observations} for signal strength measurements from the ground truth, because the word \textit{sample} refers to a pair of inputs and labels in machine learning.}) is often easier than obtaining a precise and up-to-date environment model.
These observations provide reliable but spatially sparse information about the true propagation conditions.
This motivates treating environment information as an uncertain prior and sparse observations as anchors to the ground truth.

In contrast to previous robustness studies based on generic stochastic noise or label corruption, we consider task-specific, physically motivated perturbations (we use the terms \textit{perturbations} and \textit{noise} interchangeably) that reflect realistic sources of error in indoor environments.
Furthermore, we study different environment encodings corresponding to information obtainable from practical sensing pipelines: binary occupancy from point clouds, semantic classes derived from images, and explicit material properties as an idealized baseline.
We investigate how much detail in the environment representation is beneficial under sparse observations, and whether training with such task-specific perturbations can improve the robustness of CNN-based radio map prediction to inaccurate environment information.

Our contributions are as follows:

\begin{itemize}
    \item We provide a synthetic indoor room dataset with realistic room layouts, furniture, and controlled perturbations in object positions, transmitter locations and material properties.
    Corresponding simulated radio maps and propagation path data are also published, as well as the code used to generate environments and to train and evaluate CNN models\footnote{\url{https://github.com/fabja19/RML_indoor/}}.

    \item By comparing different environment encodings, we observe that neither explicitly provided material property information nor material classes improve radio map prediction accuracy under sparse observations on our dataset.
    A binary occupancy encoding of object presence at different heights is sufficient in our experiments.

    \item We introduce training with task-specific environment perturbations (Simulated Noise as Data Augmentation, SNDA) and show that models trained this way are significantly more robust to inaccurate environment inputs at test time.
    We further compare the impact of different perturbation types, namely transmitter and object position errors of different severity.

    \item We validate our approach on real-world indoor measurements and show improved accuracy compared to RT and other classical interpolation-based methods.
\end{itemize}

\section{System Model and Background}\label{sec:system}

\subsection{Path Loss and Problem Definition}\label{sec:system:pldef}

Following the framework established in \cite{imgopt}, we define the path loss as the ratio between received and transmitted power, $\power_{\Rx}$ and $\power_{\Tx}$, capturing the large-scale fading effects while averaging out small-scale random fluctuations due to multipath interference. 
Formally, for a transmitter-receiver pair connected by $n$ propagation paths with complex coefficients $c_p$, the path loss is given by the sum of the magnitudes 
\begin{equation}\label{eq:pl}
    \power_{\Rx} / \power_{\Tx} = \sum_{p=1}^n \expec[\left|c_p\right|^2],    
\end{equation}
where the expectation accounts for random phase variations or equivalently represents local time-frequency averaging of the instantaneous received power, see \cite{imgopt}. 
A radio map is then defined as a function $\RM: \mathbb{D} \rightarrow \mathbb{R}$ that maps each location in a set of interest $\mathbb{D} \subset \mathbb{R}^2$ to its corresponding path loss value. 
In our work, $\mathbb{D}$ corresponds to a uniform spatial grid over a rectangular area, allowing us to represent the radio map as a matrix or image where each pixel encodes the path loss at that location. 
Path loss values are typically expressed in dB scale and truncated at a noise floor threshold below which signal levels are negligible for practical applications.

We consider the supervised learning problem of predicting complete radio maps from available, potentially noisy, environmental data and partial observations. 
Given environmental geometry and transmitter information, along with a set of sparse ground truth (GT) path loss observations at specific receiver (Rx) locations, our goal is to predict the full radio map across all locations of interest. 
In the indoor scenarios considered in this work, the environmental information $X$ comprises room layouts (walls, openings), and the positions and shapes of furniture and other objects and of the transmitter. 
Let $S \subset \mathbb{D}$ denote the set of locations where sparse GT measurements are available. 
Our objective is to learn a mapping $f_\theta: (X, {\RM_{gt}(s)}_{s\in S}) \rightarrow Y$, where $Y$ represents the predicted radio map $\RM_{pred}: \mathbb{D} \rightarrow \mathbb{R}$ covering all locations in the domain $\mathbb{D}$. 
We optimize the model parameters $\theta$ by minimizing the mean squared error (\textit{MSE}) between the predicted and ground truth radio maps: 
\begin{equation}\label{eq:mse}
    L(\theta) = \expec_{(x,y)}[(\RM_{pred}(x,y) - \RM_{gt}(x,y))^2]
\end{equation}
, where the expectation is taken over the training dataset.

\subsection{Dataset}\label{sec:system:dataset}
\begin{figure}
    \includegraphics[width=\linewidth]{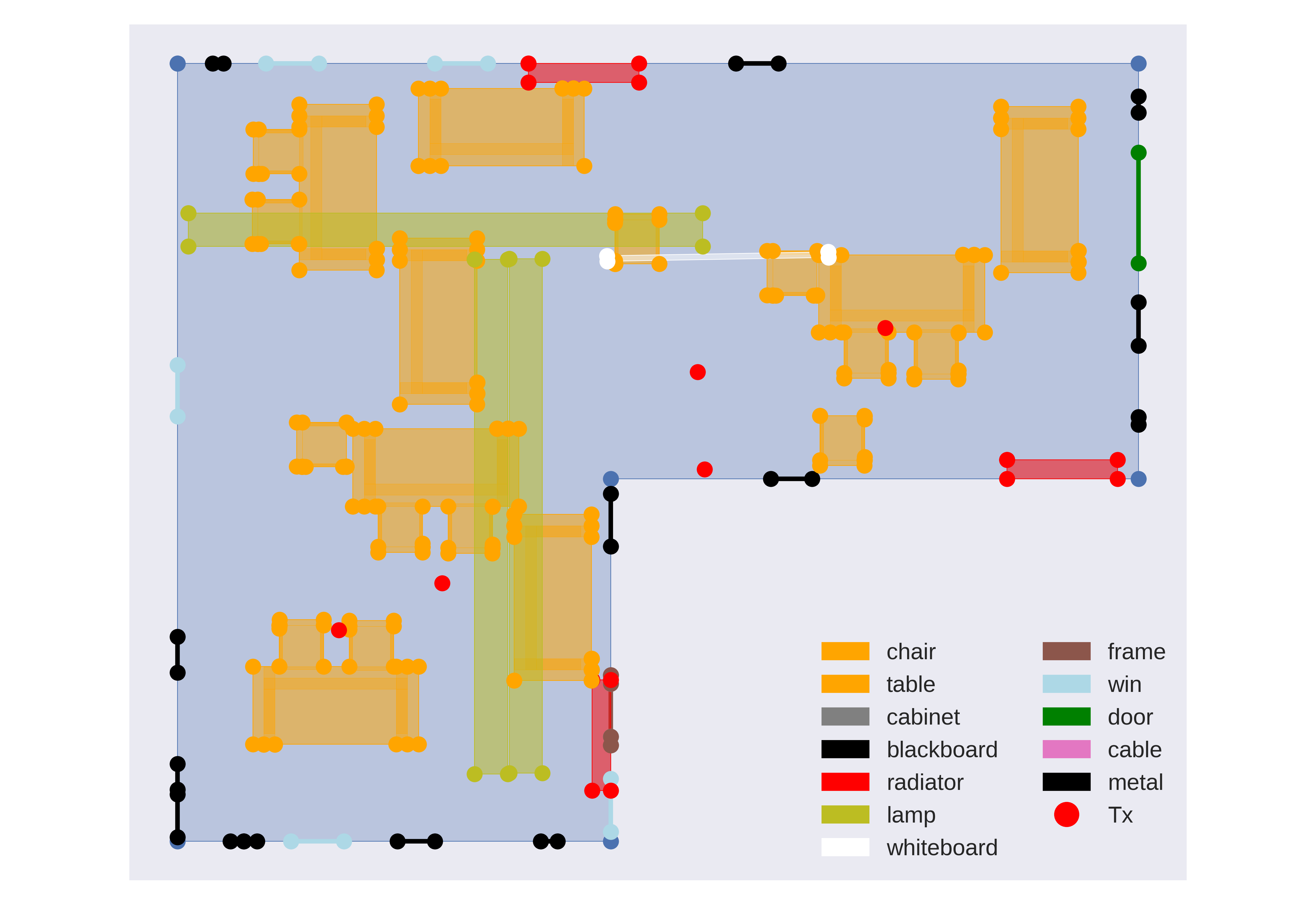} \\ 
    \includegraphics[width=\linewidth]{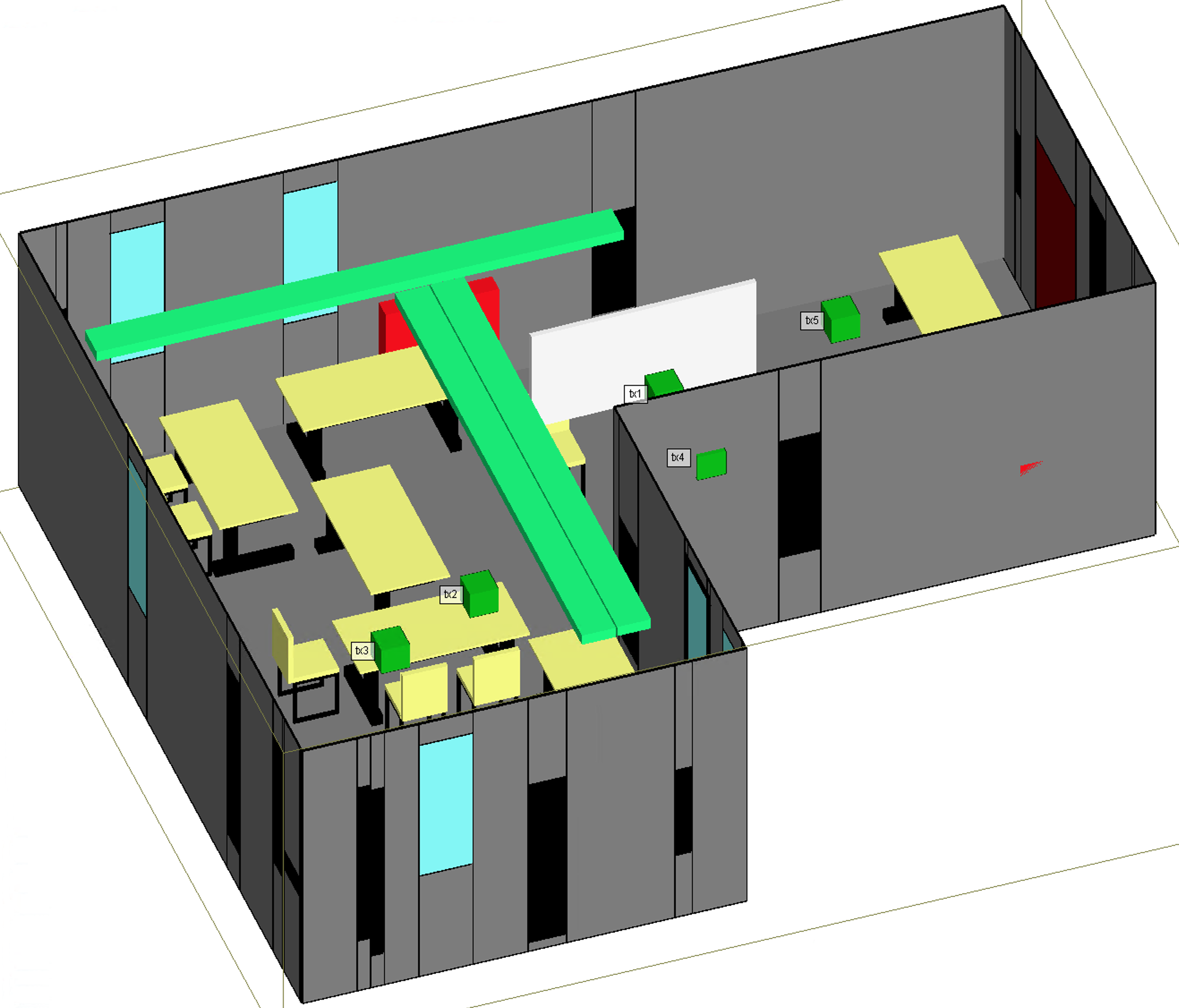} 
    \caption{Example of a room from our dataset. Plot of 2D geometry (top) and 3D model after import into ray-tracer (bottom).}
    \label{fig:project148}
\end{figure}

Due to the lack of publicly available indoor radio map datasets featuring realistic object arrangements, we developed a comprehensive dataset generation framework. 
Our codebase represents rooms and objects as polygons and lines in the x-y plane, extruded along the z-direction to form 3D structures. 
The framework enforces geometric consistency by preventing object overlaps and ensuring that all objects are positioned realistically within room boundaries. 
Each object is assigned a material type, with electromagnetic properties (permittivity, conductivity) sampled from ranges derived from the literature and ITU standards (see our codebase for specific values and sources). 
This allows automated generation of diverse environments with randomized transmitter locations, object arrangements (e.g. chairs, tables, heating radiators, lamps), and material properties.
Note that a similar approach is described in \cite{spatial_freq}, but the authors do not publish their code or the dataset.
Environments and simulation setups can be written into files readable by the RT software Wireless InSite \cite{wi}, and vice versa compatible Wireless InSite projects can be imported by our code.
See Fig. \ref{fig:project148} for an example of an environment from the dataset.

We generated 3,601 distinct base environments, placing five Txs at different locations in each. 
RT simulations using Wireless InSite \cite{wi} produce one ground truth radio map per transmitter according to \eqref{eq:pl}, yielding 18,005 samples in total. 
Simulations consider up to three reflections, three transmissions, and one diffraction per propagation path.
The carrier frequency is set to $5.92\GHz$ and Tx input power to $0\dBm$.
Rx are placed at $1$m height on a grid of $30$cm step size.
The processed radio maps have fixed pixel dimensions of $32\times 32$ pixels.
Locations outside the rooms are assigned a value of zero and are not considered in the loss calculation.
The minimum and maximum path loss in our dataset are $-71\dBm$ and $-12\dBm$, respectively, and we use a linear mapping to scale the path loss values to the range $[0,1]$ for the experiments as explained in \cite{radiounet}, \cite{imgopt}.

A key contribution of our work is the systematic investigation of training with noisy environmental inputs. 
For each base environment, we generate ten noisy copies by introducing controlled perturbations in object positions and Tx locations, with an average offset of $0.5$m, and perturbations of electromagnetic material properties. 
These perturbations simulate realistic uncertainties that arise in practice due to measurement errors, incomplete knowledge of exact object placements, or temporal changes in the environment. 
By training models on pairs of noisy inputs and corresponding ground truth radio maps, we aim to explicitly teach the model to be robust to input noise. 
We refer to this approach as \textit{Simulated Noise as Data Augmentation} (SNDA). 
As we demonstrate in Section \ref{sec:results:measurements}, SNDA not only improves model robustness at test time when faced with noisy or uncertain inputs, but also enhances generalization to real world measurements.

\subsection{Deep Learning Approach}\label{sec:system:learning}
To represent the environments in an adequate form for the CNNs, we slice and rasterize them at several heights, similar to \cite{slices}.
The values of the rasters are chosen to represent the environment in varying levels of detail:
\begin{itemize}
    \item \textit{Binary encoding:} Presence and absence of any kind of object, regardless of the material, are represented by a 1 and 0, respectively. 
        This could be retrieved from a point cloud obtained from a LiDAR scan or stereo image matching, for example.
    \item \textit{Classes:} Objects and their parts are encoded as classes corresponding to the materials (wood, metal, glass, concrete/drywall and free space). 
        This information could be obtained from images using semantic segmentation. Note that for each material, we consider a range of possible parameters as explained in Section \ref{sec:system:dataset}.
    \item \textit{Material properties:} Numeric values of electromagnetic permittivity, conductivity (in log scale) and thickness of materials are saved in the rasters. 
        This level of exact information is quite unrealistic in practice.
\end{itemize}
Tx locations are represented via one-hot-encoding showing the height value of the Tx in its location and zeros elsewhere, and a distance map containing the 2D distance in the xy-plane to each pixel location, see \cite{imgopt}.
A number of observations, i.e. samples from the ground truth radio map, are drawn randomly and provided as input in an image/tensor otherwise filled with zeros, in order to guide the prediction.
We experiment with varying resolutions of environment slices and observe that a fine pixel dimension of $256\times 256$ in the 2D plane together with slicing at $10$cm intervals in height increases accuracy compared to coarser resolutions.
The other inputs, i.e. Tx location and observation maps, are upsampled to map the resolution of the environment images.

We use the network architecture SAIPPNET from \cite{saippnet} that achieved the first rank in the sampling-assisted path loss radio map prediction competition \cite{challengeindoorsamples}.
We note that other architectures, including RadioUNet \cite{radiounet} and the UNet-based model from \cite{imgopt}, required substantially more hyperparameter tuning to achieve comparable accuracy on our dataset in preliminary experiments, and we therefore focus on SAIPPNET throughout. 
Since our main contribution lies in the training strategy (SNDA) and the analysis of environment encodings rather than in the architecture itself, we expect that our findings transfer to other CNN architectures.
We adjust the number of encoder and decoder layers to fit the resolution of inputs and outputs.

\subsection{Baseline Methods for Comparison}\label{sec:system:baselines}
As a simple baseline, we compare to free space path loss (FSPL) calculated from the carrier frequency and the distance to the Tx in each point,
\begin{equation}
    FSPL(x) = -20 \cdot \log_{10}(\|x_T - x_R\| \cdot f_c) + 147.55 + a,
\end{equation}
where $x_T, x_R$ are the 3D positions of Tx and Rx, respectively, and $f_c$ is the carrier frequency \cite[p.~48]{molisch}.
To make the comparison fairer, we add the variable term $a$ capturing additional losses or gains in the system and fit it via MSE minimization to the ground truth observations.
When comparing with the results from RT for real world measurements in Section \ref{sec:results:measurements}, we apply the same idea of fitting an additional constant to match the observations.

Furthermore, we compare to two radial basis function (RBF) methods interpolating given observations, Gaussian and Thin Plate Spline (TPS) implemented using the \texttt{scipy.interpolate.Rbf} function. 
The Gaussian RBF (later GRBF) uses an exponential kernel with a tunable shape parameter, while TPS employs a kernel that produces smooth surfaces by penalizing roughness (analogous to minimizing bending energy in physical plate models). 
These methods offer different trade-offs between flexibility and smoothness, serving as classical interpolation baselines alongside FSPL and RT for evaluating our CNN-based approach.
As an additional baseline, we also train a CNN without access to environment information, relying solely on the Tx location and given observations.

\section{Results}\label{sec:results}
\subsection{Experimental Setup}\label{sec:results:setup}
We use about 80\% of the dataset for training, 10\% for validation (learning rate reduction, early stopping and selection of the best model state) and 10\% for testing.
The training set is augmented with random flips and rotations of samples.
All models are trained with Nvidia A100 GPUs using Adam optimizer with $1e^{-4}$ initial learning rate.
The maximum number of epochs is 200, but typically training stops after 100-150 epochs due to stagnating validation loss.
We minimize MSE loss \eqref{eq:mse} calculated over the whole ground truth.

The models are trained with a random number of observations between 0\% and 20\% per iteration.
We experimented with training separate models for specific observation percentages, which does not show any benefits.
In addition, weighting the loss on the given observations higher provides no improvements.

\subsection{Numerical Results on Environment Encoding}\label{sec:results:synthetic}

\begin{figure}[h]
    \includegraphics[width=\linewidth]{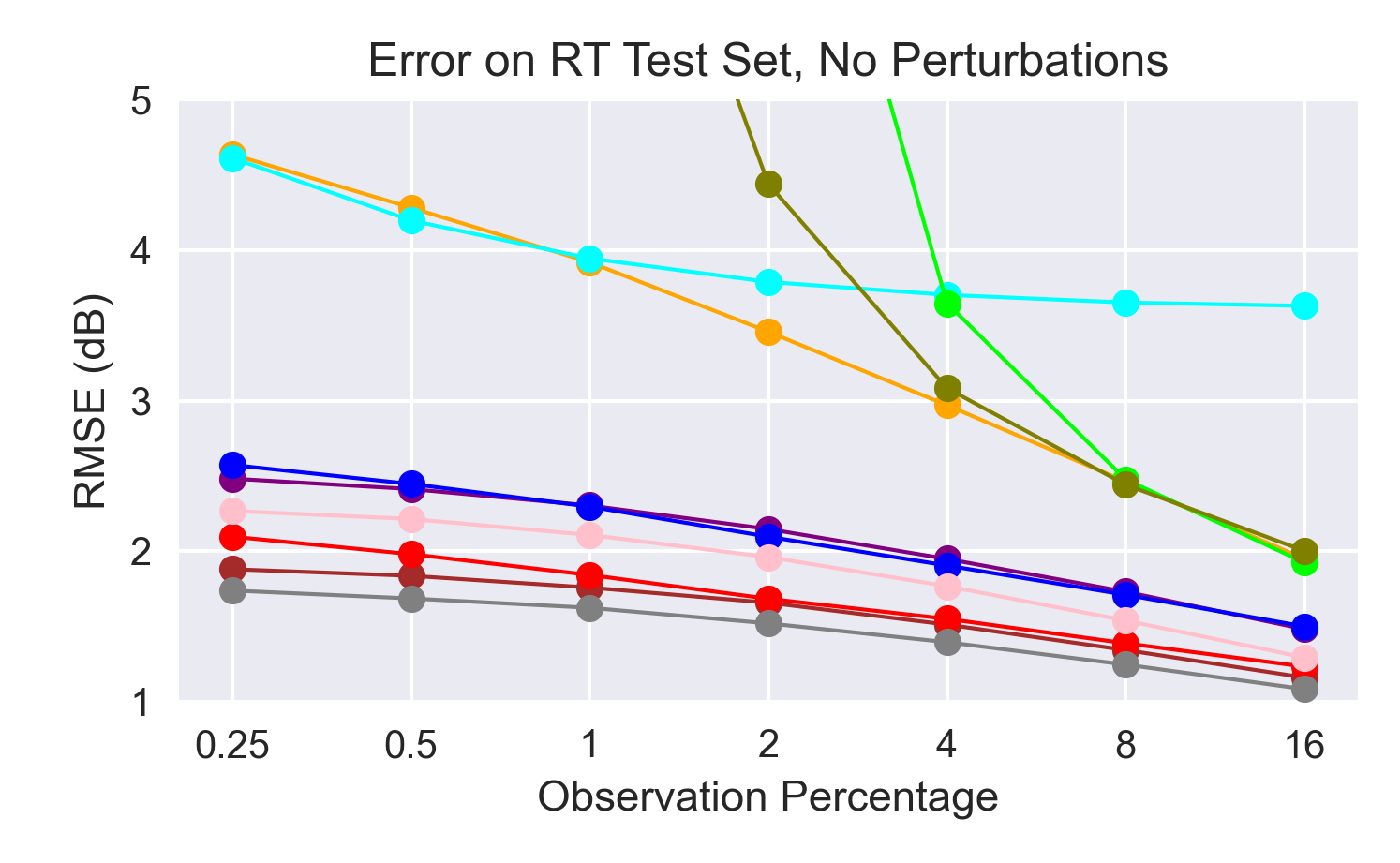}\\
    \includegraphics[width=\linewidth]{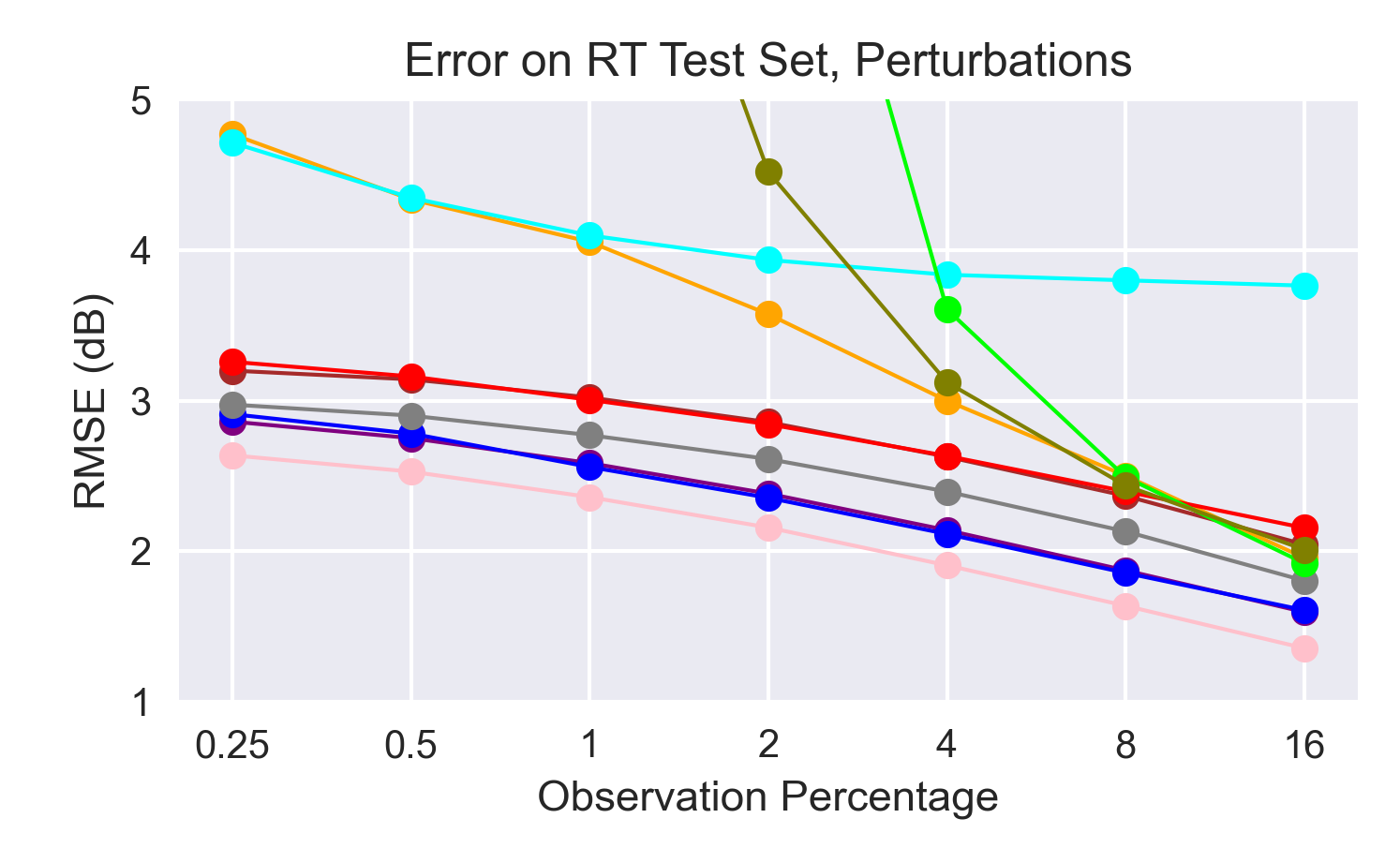}\\
    \centering
    \includegraphics[width=\linewidth]{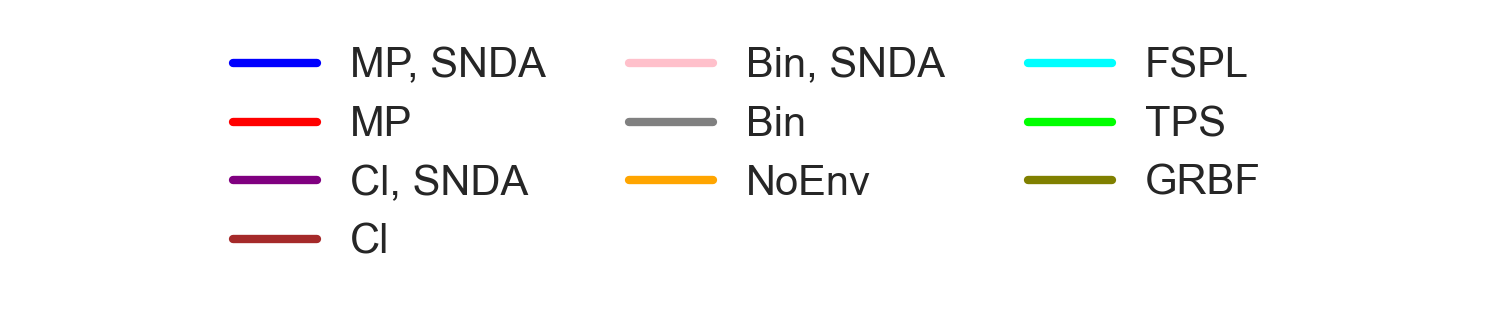}
    \caption{CNN trained with a random number of observations from $[0\%, 20\%]$ and different inputs (\textbf{M}aterial \textbf{P}roperties, \textbf{Cl}asses, \textbf{Bin}ary encoding, \textbf{No Env}ironment) compared with baseline methods on the test set generated with RT.}
    \label{fig:curves_test_rt}
\end{figure}

In Fig. \ref{fig:curves_test_rt}, we plot the performance of models trained with different environment information and the baseline methods for varying observation percentage.
We can see that up to a high observation percentage (16\%), all CNN models outperform the baselines, and the better ones even at 16\%.
The models trained with SNDA show stronger robustness when facing inaccuracies in the input information  at test time.
Also for pure interpolation (no environment information), the CNN is competitive with the classical methods.
While no environment information at all is clearly suboptimal, the results show that, within the considered encodings, the binary encoding is sufficient and more detailed classes or even material properties provide no benefit, in fact they even deteriorate performance.
This could mean that the impact of the material properties on the radio maps is rather small compared to the overall error or they may be implicitly learned from the object shapes, since our dataset contains a finite number of objects with limited variation of the electromagnetic properties per material.
For models trained without any observations, we have observed the same, which suggests that material properties are not implicitly inferred from observations.

\begin{figure}[h]
    \begin{tabularx}{.9\linewidth}{XXXXX}
        \begin{subfigure}{\imgwidth}\includegraphics[width=\textwidth]{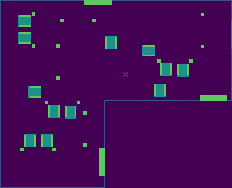}\end{subfigure} &
        \begin{subfigure}{\imgwidth}\includegraphics[width=\textwidth]{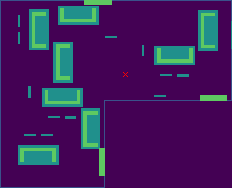}\end{subfigure} &
        \begin{subfigure}{\imgwidth}\includegraphics[width=\textwidth]{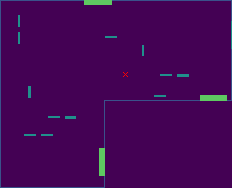}\end{subfigure} &
        \begin{subfigure}{\imgwidth}\includegraphics[width=\textwidth]{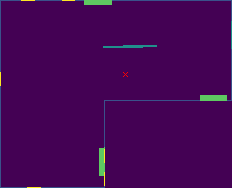}\end{subfigure} &
        \begin{subfigure}{\imgwidth}\includegraphics[width=\textwidth]{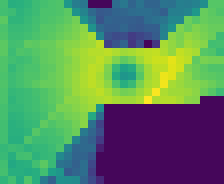}\end{subfigure} \\[1em]

        \begin{subfigure}{\imgwidth}\includegraphics[width=\textwidth]{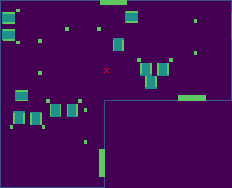}\end{subfigure} &
        \begin{subfigure}{\imgwidth}\includegraphics[width=\textwidth]{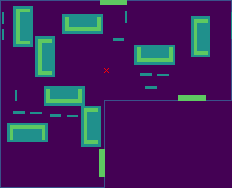}\end{subfigure} &
        \begin{subfigure}{\imgwidth}\includegraphics[width=\textwidth]{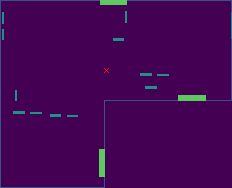}\end{subfigure} &
        \begin{subfigure}{\imgwidth}\includegraphics[width=\textwidth]{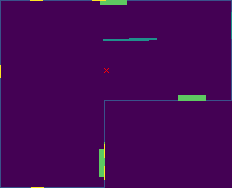}\end{subfigure} &
        \hfill\\[1em] 

        \begin{subfigure}{\imgwidth}\includegraphics[width=\textwidth]{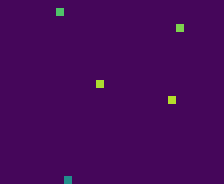}\subcaption*{1\%}\end{subfigure}&
        \begin{subfigure}{\imgwidth}\includegraphics[width=\textwidth]{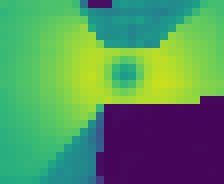}\subcaption*{2.0dB}\end{subfigure}&
        \begin{subfigure}{\imgwidth}\includegraphics[width=\textwidth]{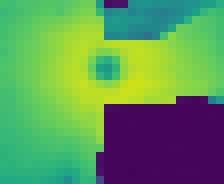}\subcaption*{4.2dB}\end{subfigure}&
        \begin{subfigure}{\imgwidth}\includegraphics[width=\textwidth]{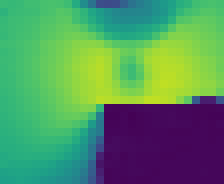}\subcaption*{2.8dB}\end{subfigure}&
        \begin{subfigure}{\imgwidth}\includegraphics[width=\textwidth]{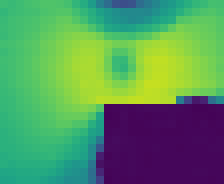}\subcaption*{3.3dB}\end{subfigure}\\
        
        \begin{subfigure}{\imgwidth}\includegraphics[width=\textwidth]{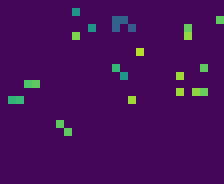}\subcaption*{5\%}\end{subfigure}&
        \begin{subfigure}{\imgwidth}\includegraphics[width=\textwidth]{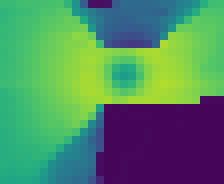}\subcaption*{1.1dB}\end{subfigure}&
        \begin{subfigure}{\imgwidth}\includegraphics[width=\textwidth]{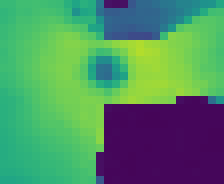}\subcaption*{3.9dB}\end{subfigure}&
        \begin{subfigure}{\imgwidth}\includegraphics[width=\textwidth]{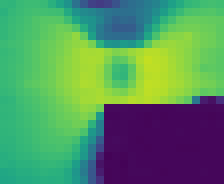}\subcaption*{2.4dB}\end{subfigure}&
        \begin{subfigure}{\imgwidth}\includegraphics[width=\textwidth]{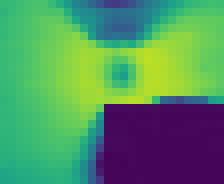}\subcaption*{2.9dB}\end{subfigure}\\
        
        \begin{subfigure}{\imgwidth}\includegraphics[width=\textwidth]{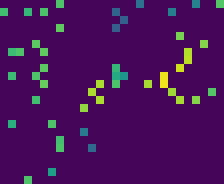}\subcaption*{10\%}\end{subfigure}&
        \begin{subfigure}{\imgwidth}\includegraphics[width=\textwidth]{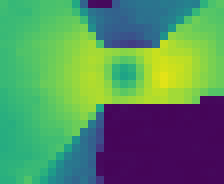}\subcaption*{1.1dB}\end{subfigure}&
        \begin{subfigure}{\imgwidth}\includegraphics[width=\textwidth]{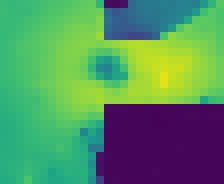}\subcaption*{3.1dB}\end{subfigure}&
        \begin{subfigure}{\imgwidth}\includegraphics[width=\textwidth]{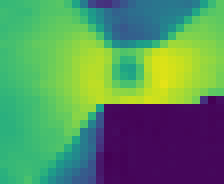}\subcaption*{1.6dB}\end{subfigure}&
        \begin{subfigure}{\imgwidth}\includegraphics[width=\textwidth]{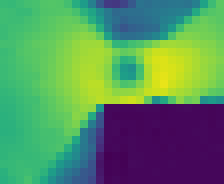}\subcaption*{2.1dB}\end{subfigure}
    \end{tabularx}
    \caption{Top row: environment slices showing material classes at 0.3/0.6/0.9/1.2m height for the environment in Fig. \ref{fig:project148}/ground truth radio map. 
    Second row: slices for noisy version of the environment. 
    Third row: 1\% Input observations and predictions for the model trained without SNDA, receiving clean environment information/ 
    the model trained without SNDA, receiving noisy environment information/ 
    the model trained with SNDA, receiving clean environment information/ 
    the model trained with SNDA, receiving noisy environment information.
    RMSE in dB below. Fourth and Fifth row: Same for 5\% and 10\% observation given. }
    \label{fig:predictions_test_rt}
\end{figure}

In Fig. \ref{fig:predictions_test_rt}, we present one of the more difficult samples from the test set.
It shows a selection of four of the slices encoding the true environment and the same slices for its noisy version, the ground truth radio map and predictions by different CNNs.
We have used the models trained on environments encoded by binary labels with and without SNDA.
In this example, the Tx offset is the primary source of inaccuracy.
The model trained without SNDA gives a pretty accurate prediction when it is given the true environment at test time as well (second column).
However, given the noisy environment, it is not capable of inferring and adjusting to the wrong information, primarily the Tx location offset (column).
The model trained with SNDA predicts more blurry shadows for low observation percentage, and it is capable of adjusting its prediction to the given observations.
Notably, even with only five observations (1\%) given, it seems to successfully correct for the wrongly given Tx location (fifth column). 
Given the clean environment, it produces more accurate results (fourth column), albeit not as precise as the model trained on clean environments.

\subsection{Sensitivity Analysis}\label{ref:results:ablation}
\begin{figure}[h]
    \includegraphics[width=\linewidth]{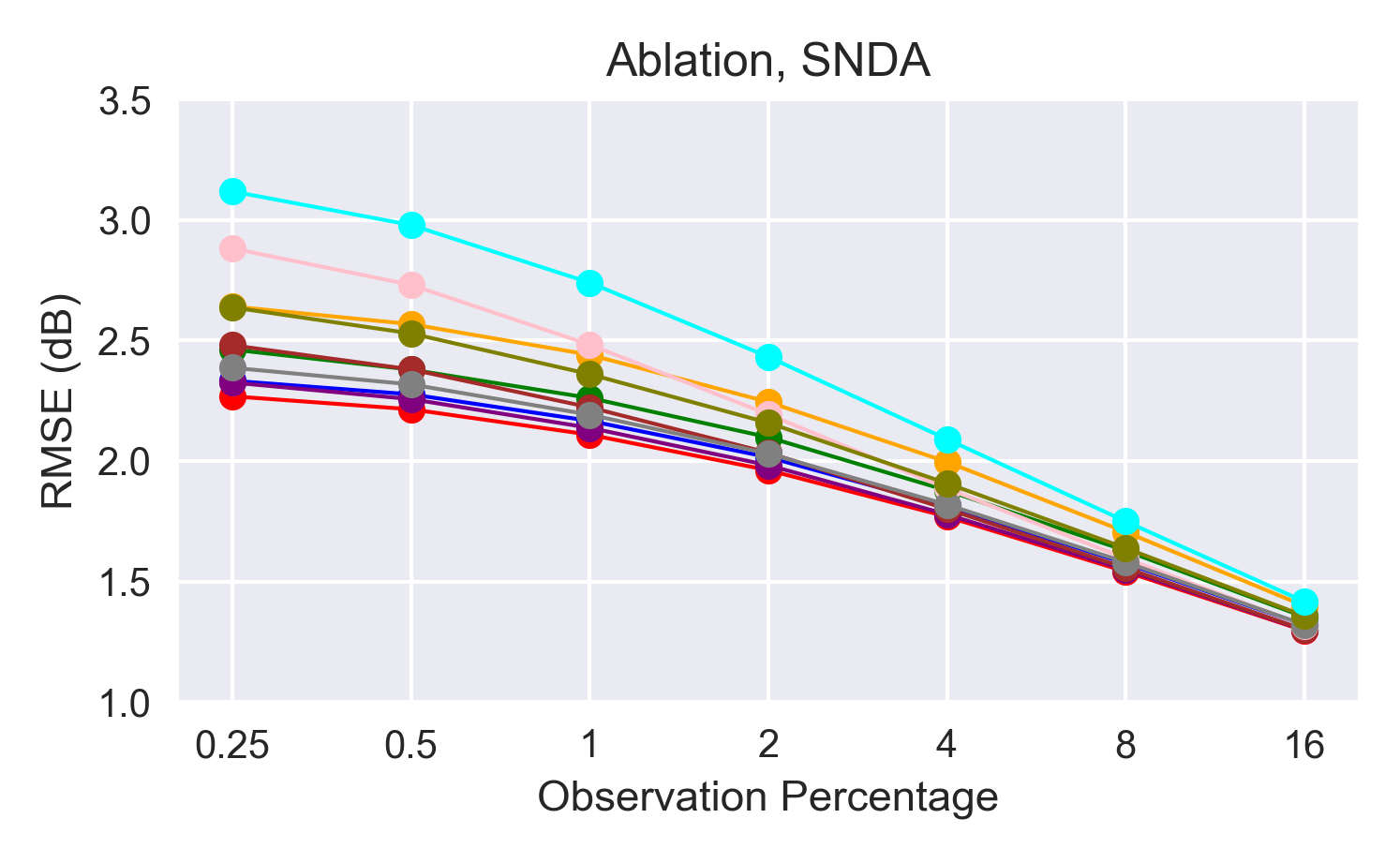}\\
    \includegraphics[width=\linewidth]{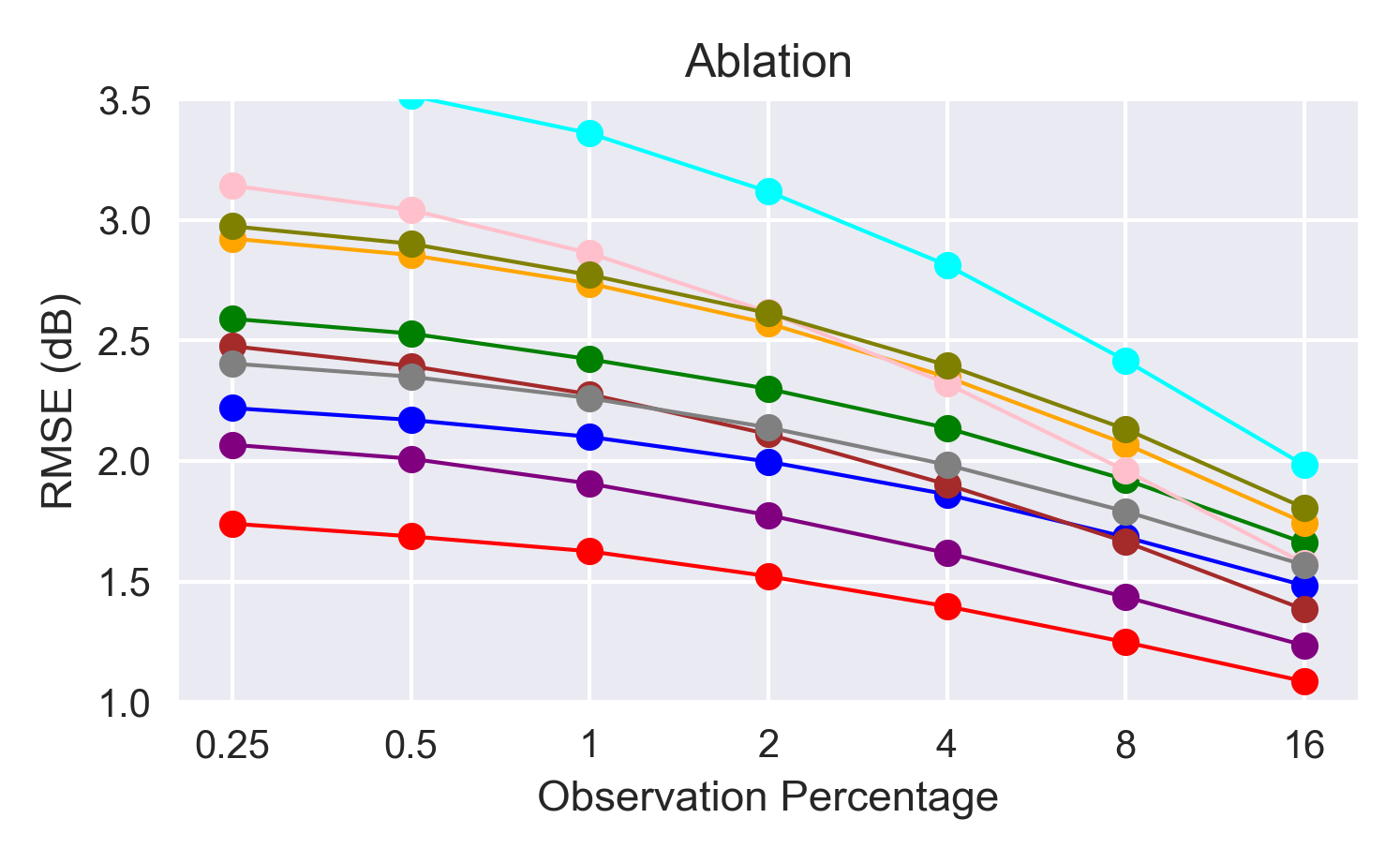}\\
    \includegraphics[width=\linewidth]{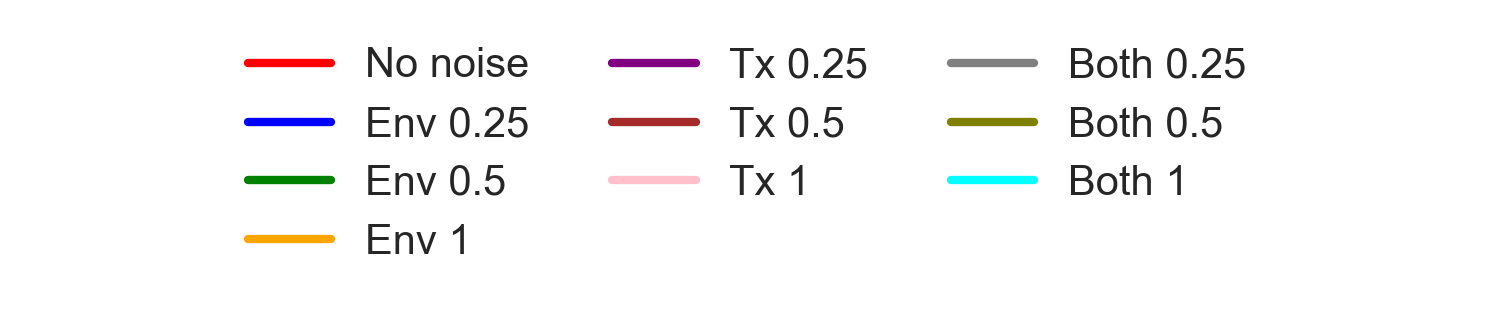}
    \caption{Test results for Perturbations of only \textbf{Tx} locations/only object locations (\textbf{Env})/\textbf{Both}/None, for varying degrees of noise and model trained with/without SNDA.}
    \label{fig:ablation_perturbations}
\end{figure}

We carry out additional tests to investigate the severity of the different kinds of noise in our data and to evaluate the robustness of the model to increased noise levels at test time.
For this, we test the model with varying degrees of offsets in the locations of Txs, objects and both together, with average offset of about $0.25$, $0.5$ and $1$m.
Recall that the results for testing with noise in Section \ref{sec:results:synthetic} correspond to an average offset of $0.5$m for both Tx and objects, which has also been used during training.
We do not test for perturbations of the material properties separately, as the best performing model used here is the one with binary inputs, that is not affected by material property perturbations.
The results are shown in Fig. \ref{fig:ablation_perturbations}.

For the model trained with SNDA, both types of separate noise have a similar impact for average offset of $0.25$ and $0.5$m.
At $1$m offset, we observe that for a low number of observations perturbation of the Tx location is more severe than object perturbations, which however changes around $1-2\%$ given observations.
We suppose that given a sufficient amount of observations, the model is able to correct the false Tx location quite accurately, which is also indicated in the example in Fig. \ref{fig:predictions_test_rt}.
For the model trained without SNDA, the results show that on noisy inputs with small deviations (i.e. offset of $0.25$m), it still performs better or on par with the model trained with SNDA.
For larger deviations, its performance degrades more severely.
We observe that, as the alignment between the training and test noise levels get closer, the performance improves, which suggests that the use of models trained under different noise levels could be beneficial.

\subsection{RESULTS ON REAL WORLD MEASUREMENTS}\label{sec:results:measurements}
\begin{figure}[h]
    \centering
    \includegraphics[height=\columnwidth,angle=270,origin=c]{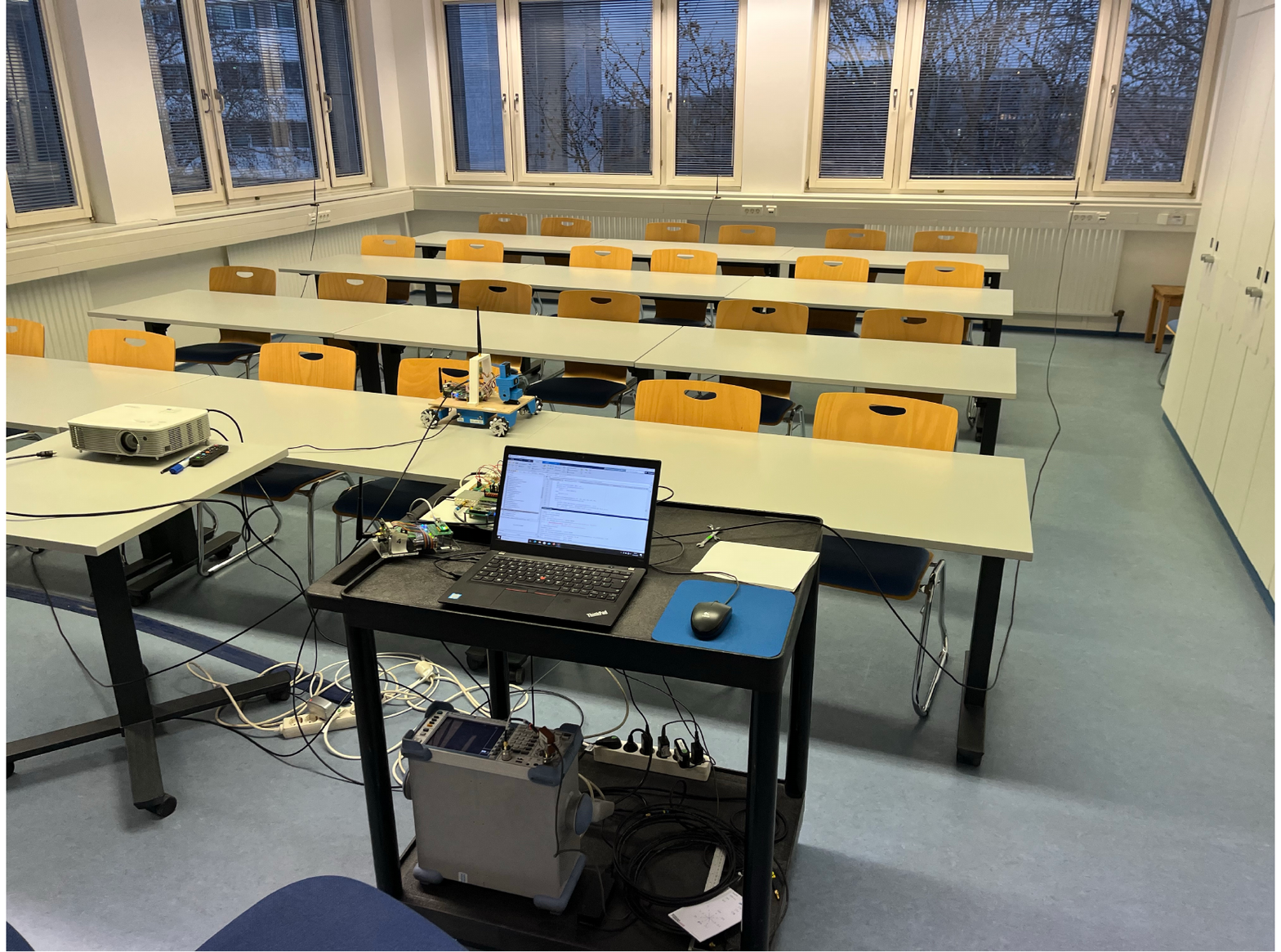}
    \includegraphics[height=\columnwidth,angle=270,origin=c]{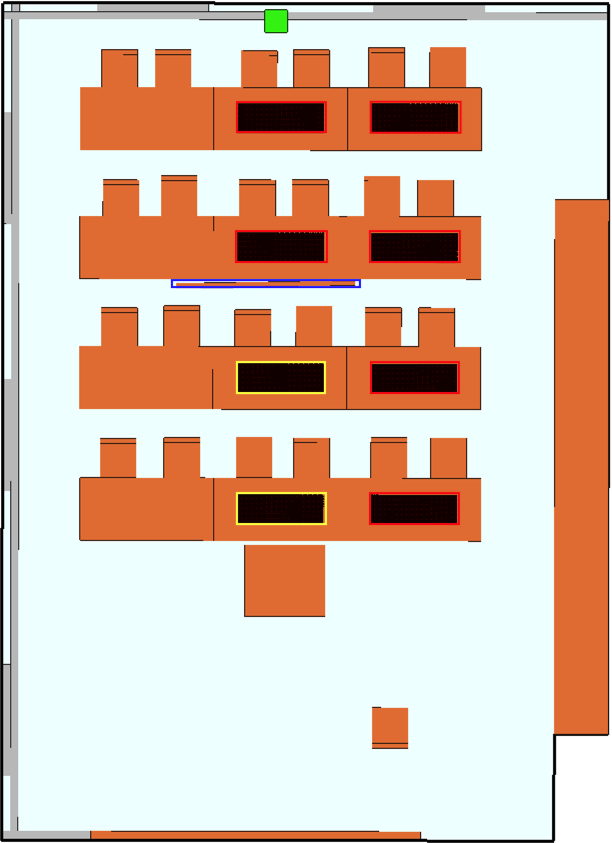}
    \caption{Measurement Scenario - photo and representation in ray-tracer with annotations: Tx in green, blockage in blue, measurement areas in black with red or yellow frame depending on whether they lie in LoS of the Tx or not, respectively.}
    \label{fig:seminar_room_setup}
\end{figure}

To evaluate whether our approach proves to be useful in real world scenarios as well, we have conducted a series of measurements in the room shown in Fig. \ref{fig:seminar_room_setup}.
A Tx (green) is placed on the edge of the room that is equipped with tables, chairs and other furniture.
Measurements are taken over several measurement areas (black with yellow/red frames) with a fine resolution of .
The path loss is defined as the local average over all measurement points belonging to a given simulation pixel, considering only pixels whose area is at least half covered by measurement points.
In our initial setup, all measurement areas were within LoS of the Tx. 
Under these conditions, the scenario is relatively simple, and even a free-space path loss (FSPL) model achieves competitive performance.
To evaluate the methods under more challenging and practically relevant conditions including non-LoS and a dynamic change in the environment that is not represented in the model input, we modify the scenario by introducing a whiteboard (blue) after taking the measurements in the red LoS areas.
This object blocks the LoS paths to the yellow areas.
We use one pixel from each of the red areas together with one pixel from each yellow area as observation input.
Additionally, the environment information either lacking or including the new object is given.
The models are tasked to predict the path loss values in the remaining pixels of the yellow areas, see Fig. \ref{fig:seminar_room_setup}.
The RMSE is calculated over 500 random draws of observation pixels following this scheme.

\begin{figure}[h]
    \begin{tabular}{@{}cc@{\hspace{0.5em}}c@{}}
        \includegraphics[interpolate=false,width=.42\linewidth]{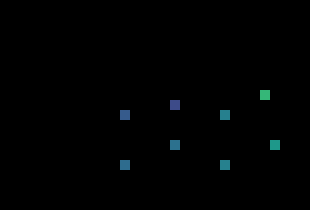} &
        \includegraphics[interpolate=false,width=.42\linewidth]{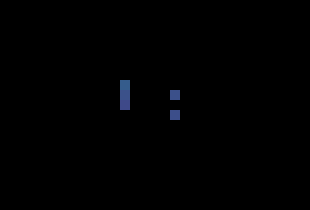} &
        \multirow{3}{*}{\includegraphics[width=0.09\linewidth]{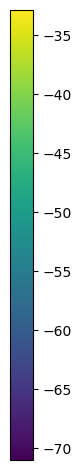}} \\
        \includegraphics[interpolate=false,width=.42\linewidth]{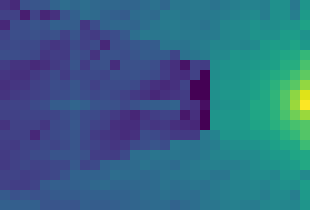} & 
        \includegraphics[interpolate=false,width=.42\linewidth]{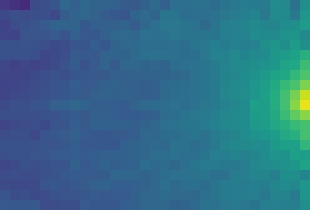} & \\
        \includegraphics[interpolate=false,width=.42\linewidth]{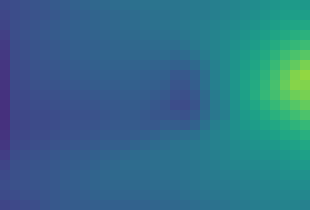} &
        \includegraphics[interpolate=false,width=.42\linewidth]{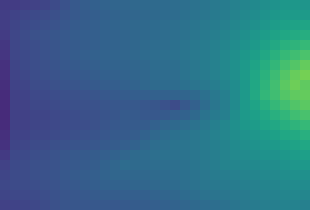} &
    \end{tabular}
    \caption{Test with measured data. 
        Top row: observations given as input and remaining pixels in nLoS-areas used as ground truth. 
        Second row: predicted radio map by ray-tracer with and without information about the object. 
        Third row: predicted radio map in dB by CNN with and without information about the object.}
    \label{fig:seminar_room_experiments}
\end{figure}

In Table \ref{tab:test_measurements}, we compare the performance of different methods. 
Our proposed CNN trained with SNDA achieves the highest accuracy in the described scenario, followed by GRBF interpolation.
We used the binary environment encoding since it provided the highest accuracy again.
If the blocking object is included in the environment information, the error is further decreased from 2.1dB to 1.3dB.
In Fig. \ref{fig:seminar_room_experiments}, we show the given observations, remaining ground truth pixels and predicted radio maps by RT and the CNN with and without information about the object.
The output confirms that the model is able to predict an estimate of the shadow in the correct region based on the given observations, even in absence of the object in the environment input.
While validation on additional rooms and transmitter configurations would further strengthen these findings, even this single scenario demonstrates the practical viability of the approach under conditions that include an unmodeled dynamic change in the environment.

\begin{table}
    \centering
    \begin{tabular}{|c|c|}
        \toprule
        Method  &   RMSE (dB)   \\
        \midrule
        CNN, Bin &   2.1 \\
        CNN, Bin, object &   1.3 \\
        CNN, NoEnv  &   1.7 \\
        RT  &   3.1 \\
        RT with object & 3.1 \\
        FSPL    &   2.8 \\
        GRBF &   2.3 \\
        TPS &   2.8 \\
        \bottomrule
    \end{tabular}
    \caption{Test error on measured data. 
            For CNN and RT "with object", the LoS blocking object is shown in the input environment, otherwise it is not. 
            The CNN with binary environment information was trained with SNDA.}
    \label{tab:test_measurements}
\end{table}

\section{Conclusion}\label{sec:conclusion}
We have investigated CNN-based radio map prediction for indoor environments under realistic conditions where environment information is incomplete and inaccurate. 
While most existing approaches assume precise environment models, we have explicitly addressed structured inaccuracies in geometry, material properties, and transmitter positions that naturally arise from sensor limitations, mapping errors, and temporal changes.

Our results demonstrate that training CNNs with task-specific perturbations significantly improves robustness to noisy inputs at test time as well. 
Models trained in this way reduce prediction error by up to 25\% compared to those trained on clean data when tested with perturbed environment information. 
This improvement is achieved by explicitly teaching the network to compensate for systematic input errors during training. 
Notably, our approach works even with very sparse observations: with as few as 1\% of locations sampled, models trained with SNDA can implicitly correct substantial errors including transmitter location offsets of around 0.5m.

A surprising finding is that binary occupancy encoding outperforms more detailed representations in our scenarios. 
Explicitly providing material properties or semantic class labels does not improve accuracy and can even slightly reduce performance. 
We suspect that this occurs because either propagation in single-room scenarios is dominated by geometric effects, or they learn typical material-shape associations from training. 
This considerably  simplifies practical deployment in such scenarios, as binary occupancy is readily obtainable from LiDAR or stereo vision without requiring material classification.

We validated our approach on real-world measurements, achieving 2.1 dB RMSE compared to 3.1 dB for RT, 2.8 dB for FSPL and TPS and 2.3dB for GRBF. 
When the line-of-sight blocking object is included in the environment representation, the error further decreases to 1.3 dB. 
This demonstrates that CNN-based prediction trained under realistic uncertainty can serve as a practical alternative to computationally expensive ray-tracing.

Our study focuses on single-room scenarios with furniture and other objects and at most two transmissions through walls.
Most other works, except \cite{spatial_freq} have considered environments with multiple rooms but no objects. 
Multi-room propagation involving a larger number of wall penetration could potentially show different behavior and may benefit more from material information. 
As a future work, it would be interesting to investigate the performance in scenarios with multiple rooms and furniture.
Additionally, our experiments are conducted at a single carrier frequency of 5.92 GHz. 
Investigating the generalization of our findings to other frequency bands, particularly millimeter-wave frequencies where material properties are reported to play a larger role in propagation, is an interesting direction for future work.
 
\section{Acknowledgments}
We would like to thank Andreas Kortke for the measurement data and Tom Burgert for advice on label noise and sensitivity analysis.

\bibliographystyle{IEEEtran}
\bibliography{bib_indoor}

\end{document}